# Orbit Tracking Control of Quantum Systems

Shuang Cong[1)2)] Jianxiu Liu[1)], Fei Yang[1)]

[1]Department of Automation, University of Science and Technology of China, Hefei, 230027, China

[2]Key Laboratory of Quantum Information, USTC, Chinese Academy of Sciences, Hefei 230026, China

**Abstract:** The orbit tracking of free-evolutionary target system in closed quantum systems is studied in this paper. Based on the concept of system control theory, the unitary transformation is applied to change the time-dependent target function into a stationary target state so that the orbit tracking problem is changed into the state transfer one. A Lyapunov function with virtual mechanical quantity *P* is employed to design a control law for such a state transferring. The target states in density matrix are grouped into two classes: diagonal and non-diagonal. The specific convergent conditions for target state of diagonal mixed-states are derived. In the case that the target state is a non-diagonal superposition state, we propose a non-diagonal *P* construction method; if the target state is a non-diagonal mixed-state we use a unitary transformation to change it into a diagonal state and design a diagonal *P*. In such a way, the orbit tracking problem with arbitrary initial state is properly solved. The explicit expressions of *P* are derived to obtain a convergent control law. At last, the system simulation experiments are performed on a two-level quantum system and the tracking process is illustrated on the Bloch sphere.

## 1 Introduction

Advance in laser technology has prompted the development of quantum control. Over the last few decades, a series of research results have been widely used in chemical reactions [1], molecular motion [2], quantum teleportation [3] and so on. After the intensive study for half a century, many methods in classical control theory have been introduced into microcosm conditionally, such as optimal control [4-9], adaptive control [10-12] and the Lyapunov-based control [13-19, 22]. According to the fact that the system is isolated or interacts with environment, a quantum system can be divided into the closed or open quantum systems. In a closed quantum system, the system is in unitary evolution. In an open quantum system, the leakage of information from system to environment leads to the un-unitary evolution. The commonly used master equation for an open quantum system is in Lindblad style: $\dot{\rho}=-\frac{i}{h}[H_0,\rho]+L_D(\rho)$, which consists of a closed system $\dot{\rho}=-\frac{i}{h}[H_0,\rho]$ with a dissipation item $L_D(\rho)$. Obviously, the feature analysis of closed quantum systems is easier. In fact, the study of closed quantum systems is the basis of open quantum systems' research. Currently, there are still a lot of unresolved problems in control characteristics and methods in closed quantum systems. They deserve to be further studied.

From the system control perspective, two types of control problems can be classified in quantum systems. One is the state transferring or state preparation, which means to transfer an arbitrary initial state to a desired target state under the action of control law designed by means of a suitable kind of system control theory. The other one is the tracking control which consists of the orbit tracking and the trajectory tracking. The orbit tracking refers to track a free-evolutionary system of the target system, where the target system has the same free Hamiltonian $H_0$ with the system controlled. The orbit tracking is a unique phenomenon in quantum systems because there always exists the free-evolution even without control. The trajectory tracking contains two kinds. The first tracking is to track another target quantum system: $i\hbar \frac{\partial}{\partial t}\hat{\rho}_f(t) = \left[ H_{0f} + \sum_r f_r H_r, \hat{\rho}_f(t) \right]$, which may possess the same free Hamiltonian with the system controlled or the different one. In this case, it is just the orbit tracking when $H_{0f}=H_0$ and $f_r=0$. So the orbit tracking is a special case of the trajectory tracking. The second kind is to track an arbitrary time-dependent function, such as ramp function, step function and so on. The trajectory tracking problem is usually solved by way of changing it into the state transferring (steering) problem. So the key point of solving the trajectory tracking problem in quantum systems is to solve the problem of state transferring, which has been a research hotspot for a long time [5, 6, 20]. In recent years, some papers on quantum orbit tracking were also reported [7, 8, 13, 14, 19, 21]. Among them, optimal control was employed to study orbit tracking [7, 8], where the orbit tracking of the target state with non-degenerate eigen-spectrum was fully studied [13]. Furthermore, the convergent control laws for pure states were investigated [19,21]. Among many control methods, the control strategy based on Lyapunov stability theorem is local optimal. It is only a stable control in general, which has an error in the control system. Because the control in quantum systems is probabilistic, it can not guarantee the system converges to the desired target state under a stable control law if there is an error. For this reason, a convergent rather than just stable Lyapunov control law is needed in quantum systems control, which has been a research focus for several years. So far, the effective control laws based on Lyapunov method for convergent state transferring with target state of eigenstate and diagonal mixed-state in closed quantum system were solved [20,24], however, it is still an open problem to completely transfer the non-diagonal target states. A Lyapunov function based on virtual mechanical quantity $P$ was proposed to get the convergent conditions of the diagonal mixed-state [24]. However, it did not give specific instructions on how to design the $P$.

The orbit tracking of free-evolutionary target system in quantum systems is studied based on the Lyapunov method in this paper. After changing the tracking problem into the state transfer problem, we focus on the convergence problem of state transferring. Two objectives are expected: 1) a convergent control law is derived for complete state transfer between arbitrary states; 2) a control method for tracking a free-evolutionary target system is proposed based on 1).

The rest of the paper is organized as follows. In Sec.2 the system model is described by the Liouville equation. The Lyapunov stability theorem is used to design the orbit tracking control law in Sec. 3. Sec. 4 is divided into two parts, the first one 4.1 is to handle the convergence of initial target state with diagonal mixes-state and the second part 4.2 is the one for non-diagonal initial target state. In Sec.5, numerical simulation experiments are performed on a two-level system to demonstrate the effectiveness of the control strategy. Finally, Sec.6 concludes this paper.

## 2  Description of the control system model

The control system model studied in this paper is described as:

$$i\hbar \frac{\partial}{\partial t}\hat{\rho}(t) = \left[H_0 + \sum_{m=1}^{M} f_m(t) H_m, \hat{\rho}(t)\right] \qquad \hat{\rho}(0) = \hat{\rho}_0 \tag{1a}$$

$$i\hbar \frac{\partial}{\partial t}\hat{\rho}_f(t) = \left[H_0, \hat{\rho}_f(t)\right] \qquad \hat{\rho}_f(0) = \hat{\rho}_{f0} \tag{1b}$$

where $H_0$ is the free Hamiltonian representing the energy of system (1). $H_m$ represent system's control Hamiltonians, all of them will be assumed to be time-independent. $f_m(t)$ are time-dependent external control fields. The Planck constant is chosen as $\hbar = 1$ for convenience.

For the orbit tracking problem in system (1), the target system (1b) is a time-dependent quantum system. To deal with the problem, a unitary transformation $U(t) = \exp(-itH_0)$ is introduced, and states in (1) become $\rho(t) = U^\dagger(t)\hat{\rho}(t)U(t)$, $\rho_f(t) = U^\dagger(t)\hat{\rho}_f(t)U(t)$, where "†" denotes conjugate, "^" denotes states before unitary transformation. After this transformation the system (1) is represented by:

$$i\frac{\partial}{\partial t}\rho(t) = [\sum_{m=1}^{M} f_m(t) H_m(t), \rho(t)] \quad \rho(0) = \hat{\rho}_0 \tag{2a}$$

$$i\frac{\partial}{\partial t}\rho_f(t) = 0 \qquad \rho_f(0) = \hat{\rho}_{f0} \tag{2b}$$

where $H_m(t) = U^\dagger(t) H_m U(t)$.

It can be seen that the dynamic system (2a) is governed by the new Hamiltonian $H_m(t) = e^{iH_0 t} H_m e^{-iH_0 t}$ after the transformation. The control systems are changed from the Schrodinger picture into the interaction picture, and target system (1b) becomes Eq. (2b) which means target state $\rho_f(t)$ is equal to a constant, because of $\rho_f(0) = \hat{\rho}_{f0}$, we have:

$$\rho_f(t) = \hat{\rho}_{f0} \tag{3}$$

Eq. (3) means that the target state (3) in (2b) is the initial state $\hat{\rho}_{f0}$ of the original target system (1b), and the tracking problem is changed into the state $\rho(t)$ transfer problem after the transformation. In the remainder of this paper, we replace the initial state of the target system with target state for convenience.

Comparing system (1) with (2a) and (3), one can see that:
1) The system (1) is autonomous but the (2) is not.

2) After the unitary transformation, the control Hamiltonian $H_m$ in system (1), which is time-independent, becomes $H_m(t) = e^{iH_0 t} H_m e^{-iH_0 t}$ in (2) with time-dependent.

3) The time-dependent target state $\rho_f(t)$ in (1b) becomes a stationary state $\hat{\rho}_{f0}$ in (3).

Now, the control objective becomes to design a convergent control law to steer state $\rho(t)$ of (2a) to target state $\hat{\rho}_f(0)$ of (3). Since the evolution of a closed quantum system is unitary, the spectrum of $\rho(t)$ is time-invariant during the whole evolution, viz. $Tr[\rho^n(t)] = Tr[\rho_f^n(t)]$.

## 3 Control Law Derivation

Among many control methods, the Lyapunov method is simpler and easy to design. The basic idea of Lyapunov method is to select a Lyapunov function $V(x)$ satisfied the following three conditions: a) $V(x)$ is continuous and its first-order partial derivatives is also continuous in its definition; b) $V(x)$ is positive semi-definite, i.e., $V(x) \geq 0$; c) The first order time derivative of the Lyapunov function is negative semi-definite, i.e., $\dot{V}(x) \leq 0$.

There are three kinds of Lyapunov functions often used [20]. Here the Lyapunov function based on virtual physical quantity $P$ is chosen:

$$V(\rho) = tr(P\rho) \tag{4}$$

where $P$ is the observable operator, which is also called a virtual mechanical quantity operator.

To obtain a stable control law, the first-order time derivation of function (4) is obtained as:

$$\dot{V} = -\sum_{m=1}^{M} f_m(t) tr\left(iH_m(t)[\rho(t), P]\right) \tag{5}$$

For the sake of simplicity and availability, we let each item on the right side of (5) of summation sign be non-positive in order to ensure $\dot{V} \leq 0$. The control law can be derived as:

$$f_m(t) = -k_m tr\left(iH_m(t)\big([\rho(t), P]\big)\right), \quad k_m > 0 \tag{6}$$

where, $k_m > 0$ is the control gain to adjust the convergence speed of the system state.

The control law (6) designed by Lyapunov stability theorem with Lyapunov function (4) will realize the orbit tracking of a free-evolutionary target system (1b). However, control law (6) is only a stable control, which can not guarantee that the system will converge to target state. For this reason, we need to do further study to get the convergence conditions, which will guide people to design a convergent control law. Next, we study this problem in detail.

## 4 Convergence analysis

In Sec. 3 we complete the task of changing the orbit tracking into the state transferring by means of the unitary transformation, and the desired target becomes a time-invariant state $\hat{\rho}_{f0}$. In fact the variety of quantum states, such as eigenstate, superposition state and mixed-state, produces the different forms of target state. Generally, the target states in density matrix can be divided into two kinds. The first one is diagonal target states including eigenstates and some mixed-states, which can be represented by a diagonal density matrix. So far, the convergence analysis of eigenstates has been well solved [13, 20]. Some research results have also been obtained for diagonal mixed-states [24]. The second kind is non-diagonal target states including superposition state and some mixed-states. It is so far unresolved fully.

In the existing research results, for autonomous system (1), the LaSalle's invariant principle can be used to analyze convergence, where two assumptions are needed [23]: Assumption 1: $H_0$ is strongly regular, i.e., all the transition frequencies (differences of pairs of energy levels) are different, viz. $\Delta_{jk} \neq \Delta_{pq}$, $(j,k) \neq (p,q)$, where $\Delta_{jk} = \lambda_j - \lambda_k$ and $\lambda_j$ is an eigenvalues of $H_0$. Assumption 2: Control Hamiltonian $H_m$ is fully connected: $H_m \in \{\hbar h_{jk} \mid h_{jk} = |j\rangle\langle k| + |k\rangle\langle j|, j > k\}$, where $|j\rangle$ is the eigenstate associated with $\lambda_j$. Because of the unitary evolution of closed quantum system, if the target state is reachable, which must be unitarily equivalent to the initial state, i.e., there exists a unitary transformation $U$ such that $\hat{\rho}_0 = U\hat{\rho}_{f0}U^\dagger$. We make the $\hat{\rho}_0 = U\hat{\rho}_{f0}U^\dagger$ as Assumption 3. However, on the one hand, the LaSalle's invariant principle fails to deal with the situation in this paper because the system (2) is a non-autonomous system. On the other hand, based on the above three assumptions, the "Lyapunov-like lemma", which is also called improved Barbalat lemma, can be applied to the non-autonomous system with the following contents [25]: If scalar function $V(x,t)$ satisfies: (1) $V(x,t)$ is lower bounded; (2) $\dot{V}(x,t)$ is negative semi-definite; (3) $\dot{V}(x,t)$ is uniformly continuous in time. Then, $\dot{V}(x,t) \to 0$ as $t \to \infty$. By selecting the Lyapunov function (4) in this paper, one can find that all the three conditions of Lyapunov-like lemma mentioned above are satisfied: (1) $V = tr(P\rho) \geq 0$ is lower bounded for a positive $P$; (2) Its first order derivative of $V(x,t)$ is negative semi-definite under control law (6); (3) The third condition can be replaced by the existence and continuity of the second derivation of $V(x,t)$. In our paper, $\ddot{V}(\rho,t) = -\sum_{m=1}^{M} f_m(t)\{tr(i\dot{H}_m(t)[\rho,P]) + tr(iH_m(t)[\dot{\rho},P])\}$ is bounded for a bounded input.

According to the Lyapunov-like Lemma, the first derivation of the Lyapunov function (4) converges to zero for $t \to \infty$, viz., $\dot{V}(\rho(\infty),\infty) = 0$. The trajectory of controlled system (2) under Lyapunov function (4) will converge to the limit set at $t \to \infty$, which is denoted as $\mathcal{R}_1$. The states

that make Eq. (5) be zero make up the set $\mathcal{R}_1$:

$$\mathcal{R}_1 \equiv \{\rho_s : \dot{v}=0\} = \{\rho_s : tr(iH_m(t)[\rho_s, P]) = 0, \forall m, t\} \quad (7)$$

where, $\rho_s$ denote critical stable points of system (2), and $\rho_f \in \mathcal{R}_1$.

According to the formula of control law (6), the states in $\mathcal{R}_1$ satisfy $f = 0$. For the non-autonomous system (2), if $\rho \in \mathcal{R}_1$, then $\dot{\rho}=0$ holds for $f = 0$. To increase the convergence probability of target state, one way is to shrink this limit set $\mathcal{R}_1$. And then Proposition 1 is obtained:

**Proposition 1:** According to Assumption 1 and 2, the limit set $\mathcal{R}_1$ in (7) can be redefined as: $\mathcal{R}_1 \equiv \{\rho_s : [\rho_s, P] = D\}$, where $D$ is a diagonal matrix.

The proof of proposition 1 is seen in Appendix 1.

There are two significant points:

1) If $P$ is chosen as a diagonal matrix, then limit set is reduced to $\mathcal{R}_1 \equiv \{\rho_s : [\rho_s, P] = 0\}$.

2) If the case 1) is not true, one denotes $[\rho, P] = Ad_p(\vec{\rho})$, where $Ad_p$ is a linear map from Hermitian or anti-Hermitian matrices into $su(n)$. Let $A(\vec{P})$ be the real $(n^2-1)*(n^2-1)$ matrix corresponding to the Bloch representation of $Ad_p$. Denote $su(n) = T \oplus C$ and $R^{n^2-1} = S_T \oplus S_C$, where $S_C$ and $S_T$ are real subspaces corresponding to the Cartan and non-Cartan subspaces, $C$ and $T$, respectively. Let $\tilde{A}(\vec{P})$ be the first $n^2-n$ rows of $A(\vec{P})$. Then the following lemma 1 is obtained: given a generic $P$, if $rank\tilde{A}(\vec{P}) = n^2-n$ holds, then the limit set $\mathcal{R}_1$ is regular, viz. $\mathcal{R}_1 \equiv \{\rho_s : [\rho_s, P] = 0\}$. (It can be seen from the proof of Lemma V.4 in [13]).

Based on the above analysis, we would choose $P$ with different eigenvalues and satisfy the condition of lemma 1. Then it means that the critical points $\rho_s$ of (2) and $P$ are commutative, and the limit set $\mathcal{R}_1$ is reduced as:

$$\mathcal{R}_2 \equiv \{\rho_s : [\rho_s, P] = 0\} . \tag{8}$$

In the following content, it is shown that there cannot be only one option for $P$. We can always choose a suitable $P$ to get a limit set as (8).

The system (2) will converge to limit set (8). Whether the system converges to the target state depends on the relative positions among target state and controlled initial state and the other stable states. To make the system converge to target state, the following condition is needed, viz. [24]:

$$v(\rho_f) < v(\rho_0) < v(\rho_s) \tag{9}$$

There are three things needed to do according to (9): 1) the target state $\rho_f$ must make the Lyapunov function (4) be in its minimum value; 2) the initial state $\rho_0$ makes the value of (4) be sub-minimum; 3) Lyapunov function values of states in $\mathcal{R}_2$ except target one are larger than that of $\rho_0$. Now, the control law (6) holds which means $\dot{v} \leq 0$, so the monotonically decreasing function (4) evolves towards a smaller value. For this reason, if the condition (9) is satisfied, the system (2) started with initial state $\rho_0$ will uniquely converge to target state.

Eq. (9) is the condition to ensure the convergence of the controlled system. How to realize (9) is another key study in this paper. We'll focus on how to construct $P$ to meet (9) in next Section and concentrate on the instruction of structure of $P$ which is satisfied (9) for the diagonal and non-diagonal target states, respectively.

### 4.1 The case of diagonal mixed target state

The states with diagonal density matrixes include eigenstates and mixed-states. Ref [13, 20] gave the concrete convergence conditions for target state with eigenstates, and ref [24] proposed the convergence condition for diagonal mixed-states but not indicated how to design convergence parameters. On the basis of [24], we'll analyze detailedly how to construct the virtual mechanical quantity $P$ in Lyapunov function to meet condition (9).

Suppose the target state $\rho_f$ is a diagonal mixed-state and $\{\lambda_i, i = 1, 2 \cdots n\}$ is the eigen-spectrum of $\rho_0$. The target state $\rho_f$ should be a permutation of $\{\lambda_i, i = 1, 2 \cdots n\}$, viz. $\rho_f = diag(\lambda_1, \lambda_2 \cdots \lambda_n)$. According to the set $\mathcal{R}_2 \equiv \{\rho_s : [\rho_s, P] = 0\}$ in (8), a diagonal $P$ is the simplest choice. The other states $\rho_s$ in $\mathcal{R}_2$ are the different permutations of eigen-spectrum. In order to construct a $P$ satisfied (9), three steps need to be performed:

Firstly, $P$ is constructed to make $\rho_f$ be the point corresponding to the minimum of

Lyapunov function (4), which is realized by the following lemma 2:

**Lemma 2:** If the diagonal target state is $\rho_f = diag(\lambda_1, \lambda_2 \cdots \lambda_n)$, the matrix $P$ corresponding to $\rho_f$ is $P = diag(p_1, p_2, \cdots p_n)$, then $\rho_f$ is the point for Lyapunov function (4) to be minimum if the diagonal element $p_i$ of $P$ meets $(\lambda_i - \lambda_j)(p_i - p_j) < 0, \forall i \neq j$.

The proof of Lemma 2 is in Appendix 2.

Secondly, based on Lemma 2, a further study on $P$ is carried and the (9) is divided into two parts:

(1) $v(\rho_f) < v(\rho_0)$

The condition $v(\rho_f) < v(\rho_0)$ indicates that the value of function (4) on initial state is larger than that of target state. Otherwise, it is inconsistent with the monotonically decreasing of (4) and the target state will be unreachable. It is easy to obtain $v(\rho_f) - v(\rho_0) = \sum_{i=1}^{n}(P)_{ii}(\lambda_i - (\rho_0)_{ii})$. Based on the relationship between eigenvalues and matrix diagonal elements, the expression $\sum_{i=1}^{n}\lambda_i = \sum_{i=1}^{n}\mu_i = \sum_{i=1}^{n}(\rho_0)_{ii} = 1$ holds, where $(\rho_0)_{ii}$ is the $i$-th diagonal element of initial state $\rho_0$. So there must be at least one $k$ to make $\lambda_k < (\rho_0)_{kk}$ hold. If one wants to make $v(\rho_f) - v(\rho_0) = (P)_{kk}(\lambda_k - (\rho_0)_{kk}) + \sum_{i=1,i\neq k}^{n}(P)_{ii}(\lambda_i - (\rho_0)_{ii}) < 0$ hold, where $(P)_{kk}$ is the $k$-th diagonal element of $P$, then $(P)_{kk} > \sum_{i=1,i\neq k}^{n}(P)_{ii}(\lambda_i - (\rho_0)_{ii}) / ((\rho_0)_{kk} - \lambda_k)$ is obtained, viz. a certain $k$ satisfying $\lambda_k < (\rho_0)_{kk}$ is chosen, one gets

$$(P)_{kk} > \sum_{i=1,i\neq k}^{n}(P)_{ii}(\lambda_i - (\rho_0)_{ii}) / ((\rho_0)_{kk} - \lambda_k) \tag{10}$$

Maybe there are more than one $k$ to satisfy $\lambda_k < (\rho_0)_{kk}$, usually we choose the one $(P)_{kk}$ corresponding to a larger value to determine $v(\rho_f) < v(\rho_0)$, if not, slight regulation of $(P)_{kk}$ is performed.

(2) $v(\rho_0) < v(\rho_s)$

$\rho_s$ should be one of the permutations of eigen-spectrum. It gets $v(\rho_0)-v(\rho_s)=tr(P\rho_0)-tr(P\rho_s)$. According to Assumption 3, $\rho_0=U\rho_f U^\dagger$, so:

$$tr(P\rho_0)=tr(PU\rho_f U^\dagger)=tr(U^\dagger PU\rho_f)=\sum_{i=1}^{n}(P)_{ii}\sum_{j=1}^{n}(\rho_f)_{jj}(U_{ij})^2, \quad tr(P\rho_s)=\sum_{i=1}^{n}(P)_{ii}(\rho_s)_{ii}, \text{ and}$$

$$tr(P\rho_0)-tr(P\rho_s)=\sum_{i=1}^{n}(P)_{ii}\left(\sum_{j=1}^{n}(\rho_f)_{jj}(U_{ij})^2-(\rho_s)_{ii}\right).$$

Both $\rho_s$ and $\rho_f$ have the same spectrums, there must be $(\rho_f)_{kk}=(\rho_s)_{ii}$. Because of the unitary matrix $U$, there is $UU^\dagger=U^\dagger U=I$ and $\sum_{j=1}^{n}(U_{ij})^2=1$. One gets:

$$tr(P\rho_0)-tr(P\rho_s)=\sum_{i=1}^{n}(P)_{ii}\left(\sum_{j\neq k}^{n}(\rho_f)_{jj}(U_{ij})^2+(\rho_f)_{kk}\left((U_{ik})^2-1\right)\right)$$

$$=\sum_{i=1}^{n}(P)_{ii}\left(\sum_{j\neq k}^{n}(\rho_f)_{jj}(U_{ij})^2-(\rho_f)_{kk}\sum_{j\neq k}^{n}(U_{ij})^2\right)$$

$$=\sum_{i=1}^{n}(P)_{ii}\sum_{j\neq k}^{n}\left((\rho_f)_{jj}-(\rho_f)_{kk}\right)(U_{ij})^2$$

For the above equation, there is at least one $l$ to make $(\rho_f)_{ll}-(\rho_f)_{kk}<0$ hold. If one hopes $tr(P\rho_0)-tr(P\rho_s)<0$ holds, then the following expression is workable:

$$(P)_{ll}>\left(\sum_{i\neq l}^{n}(P)_{ii}\sum_{j\neq k}^{n}\left((\rho_f)_{jj}-(\rho_f)_{kk}\right)(U_{ij})^2+(P)_{ll}\sum_{j\neq k,j\neq l}^{n}\left((\rho_f)_{jj}-(\rho_f)_{kk}\right)(U_{lj})^2\right)/\left((\rho_f)_{kk}-(\rho_f)_{ll}\right) \quad (11)$$

The above process is to construct $P$ for mixed-states with diagonal density matrix. We conclude that: if the target state is of diagonal, a Hermitian and positive diagonal matrix $P$ is selected. To ensure convergence, the diagonal elements of $P$ must satisfy lemma 2 and Eqs. (10) and (11) simultaneously. In addition to this, we choose non-negative diagonal elements for $P$ to ensure its positivity.

### 4.2 The case of non-diagonal target density matrix

It is more complicated to analyze the convergence for the non-diagonal target state. The idea is as follows: the non-diagonal target state is changed into a diagonal matrix, and then $P$ is designed by using the method mentioned in 4.1. However, one needs notice that the superposition state is a kind of pure states, which can be represented by wave functions as $\rho_f=|\psi_f\rangle\langle\psi_f|$. As a result, a diagonal state conversion is not necessary for superposition state. What one needs to do is only to design a proper $P$. Next, we go onto the analysis of non-diagonal superposition state and mixed-state in detail.

(1) In the case of superposition target state

Before the analysis in depth, lemma 3 needs to be introduced. Lemma 3 [16]: For the $n$-level Hermitian matrix $A$ and $B$, if they are commutative, viz. $[A, B] = 0$, then both $A$ and $B$ owns the same eigenstates. We rewrite $P$ according to its eigen-decomposition as $P = \sum_k p_k |\psi_k\rangle\langle\psi_k|$, where $|\psi_k\rangle$ is its eigenstate and $p_k$ is eigenvalue. It's known that the target state $\rho_f$ can be described as $\rho_f = |\psi_f\rangle\langle\psi_f|$. According to $\rho_f \in \mathcal{R}_2$ and Lemma 3, $P$ can be described as:

$$P = p_1 |\psi_f\rangle\langle\psi_f| + \sum_{k=2}^{n} p_k |\psi_k\rangle\langle\psi_k|, \text{ where } |\psi_1\rangle = |\psi_f\rangle. \text{ And } \langle\psi_i|\psi_j\rangle = 0, \text{ for } i \neq j. \quad (12)$$

And $\rho_s$ should be:

$$\rho_s = \lambda_1 |\psi_f\rangle\langle\psi_f| + \sum_{k=2}^{n} \lambda_k |\psi_k\rangle\langle\psi_k| \quad \sum_{k=1}^{n} \lambda_k = 1. \quad (13)$$

It is known that the state $\rho_0$ and $\rho_s$ have the same spectrum under the unitary evolution, therefore $\rho_s$ has the same eigenvalues with $\rho_0$, so does the target state $\rho_f$.

Substituting $\rho_f = |\psi_f\rangle\langle\psi_f|$ into Eq. (13), the eigen-spectrum of $\rho_f$ is $\{1, 0, \cdots 0\}$. So for $\rho_s$, there is only one eigenvalue $\lambda_i$ to be non-zero, viz. $\rho_s = \lambda_i |\psi_i\rangle\langle\psi_i|$ ($\lambda_i = 1$).

To make the system be asymptotically stable, (9) is needed. If $\rho_j = |\psi_j\rangle\langle\psi_j|$ are denoted in (12), then:

$$\begin{cases} v(\rho_f) = tr(P\rho_f) = p_1 \\ v(\rho_0) = p_1 tr(\rho_f \rho_0) + \sum_{k=2}^{n} p_k tr(\rho_k \rho_0) \\ v(\rho_s) = p_j \quad (j \neq 1) \end{cases} \quad (14)$$

Combined Eq. (14) with Eq. (9), a suitable $P$ can be constructed to satisfy:

$$0 < p_1 < p_1 tr(\rho_f \rho_0) + \sum_{k=2}^{n} p_k tr(\rho_k \rho_0) < p_j \ (j \neq 1). \quad (15)$$

It can be seen from (15) that $P$ may not be a diagonal matrix for non-diagonal superposition target state. The $P$ constructed based on (12) and (15) guarantees the convergence of non-diagonal superposition target state, where (12) describes how to construct the eigenstate and (15) is to determine the eigenvalue. Moreover, the eigenvalue $p_1$ of $P$, whose corresponding eigenstate is target state, is the smallest one. And then one obtains $(1-tr(\rho_f\rho_0))p_1 < \sum_{k=2}^{n} p_k tr(\rho_k\rho_0)$ & $p_1 < p_k, \forall k \neq 1$.

In this paper, let $\frac{(1-tr(\rho_f\rho_0))}{tr(\rho_k\rho_0)} p_1 < p_k$ & $p_1 < p_k, \forall k \neq 1$, then one has:

$$g_k > \max\left\{\frac{(1-tr(\rho_f\rho_0))}{tr(\rho_k\rho_0)}, 1\right\}, p_k = g_k p_1 \tag{16}$$

where $g_k$ is the weight coefficient corresponding to $p_k$. Eq. (16) means that $p_k$ is proportional to the smallest eigenvalue $p_1$. It is obvious that Eq. (16) is not the only one for $P$. It is selected carefully to satisfy the condition of lemma 1 simultaneously.

(2) In the case of non-diagonal mixed-state

For system (1), suppose the initial target state $\hat{\rho}_{f0}$ is a non-diagonal mixed-state. The solution of (1b) is $\hat{\rho}_f(t) = e^{-iH_0 t}\hat{\rho}_{f0}e^{iH_0 t}$. To deal with this situation, a unitary transformation has been used to transform the tracking problem of (1) into the state transferring one of (2). We follow the idea of changing the non-diagonal $\hat{\rho}_{f0}$ into a diagonal one by another unitary transformation and then a convergent control law can be designed based on 4.1.

As mentioned in Section 2, the unitary transformation $U = \exp(i*H_0 t)$ is applied to change the tracking of the free-evolutionary target quantum system (1b) into the transferring one of a stationary target state $\hat{\rho}_{f0}$. $\hat{\rho}_{f0}$ is a Hermitian matrix, so it exists another unitary transformation $U_f$ to meet $U_f \hat{\rho}_{f0} U_f^\dagger = D_f$. After this transformation, The target state $\hat{\rho}_{f0}$ of (2) becomes the diagonal matrix $D_f$. In other words, two unitary transformations have been performed on (1), and then the a target system tracking with non-diagonal initial state is changed into the state transferring of diagonal target state $D_f$. In fact, in the case of orbit tracking with non-diagonal mixed target state, the problem can be solved by two in one unitary transformation $U_T$.

Let $U_T = U_f e^{iH_0 t}$, it is performed on system (1), viz., $\rho = U_T \hat{\rho} U_T^\dagger$, $\rho_f = U_T \hat{\rho}_f U_T^\dagger = U_T U_T^\dagger D_f U_T U_T^\dagger = D_f$, then the system (1) becomes:

$$i\hbar \frac{\partial}{\partial t}\rho(t) = \left[\sum_{m=1}^{M} f_m(t) H_{mt}, \rho(t)\right] \quad \rho(0) = U_f \hat{\rho}_0 U_f^\dagger \quad (17a)$$

$$i\hbar \frac{\partial}{\partial t}\rho_f(t) = 0 \quad \rho_f(0) = D_f \quad (17b)$$

where $H_{mt} = U_T H_m U_T^\dagger$.

After the unitary transformation $U_T$, the tracking of target system with non-diagonal initial state $\hat{\rho}_{f0}$ in (1) can be changed into the state transferring of a diagonal stationary state in (17). According to Proposition 1, the critical points set of system (17) is still $\mathcal{R}_2 \equiv \{\rho_s : [\rho_s, P] = 0\}$. The convergence analysis is the same as that in 4.1.

In conclusion, we have acquired the convergence conditions for non-diagonal target states.

## 5 Applications and experimental results analyses

The trajectory of quantum state in a two level system can be described by the point's trajectory in Bloch sphere, so in this Sec. a two level atom system controlled by a single control field is considered. The effectiveness of the proposed method will be illustrated.

The free Hamiltonian of the controlled system (1a) is $H_0 = \omega \sigma_z$ and the control Hamiltonian is $H_1 = \sigma_x$, where $\sigma_i (i = x, y, z)$ denotes Pauli matrix and $\sigma_x = [0\ 1; 1\ 0]$ $\sigma_z = [1\ 0; 0\ -1]$.

For this two-level system, we denotes $e_1 = |0\rangle, e_2 = |1\rangle$ as the basic vectors of $H_0$. According to (6), the control law is $f_1(t) = -k_1 tr(iH_1(t)[\rho(t), P])$. This example satisfies the three conditions in Section 4.

### 5.1 The case of non-diagonal superposition target state

The initial state of (1a) is $|\psi_0\rangle = \frac{1}{\sqrt{3}}|0\rangle + \sqrt{\frac{2}{3}}|1\rangle$ and the initial target state of (1b) is $|\psi_f\rangle = \frac{1}{\sqrt{8}}|0\rangle + \sqrt{\frac{7}{8}}|1\rangle$. They are both non-diagonal superposition states. The design process of a convergent control law is as follows:

(1) Construction of $P$

To construct $P$, a set of linearly independent vector $|\psi_k\rangle (k=1,2)$ is prepared. In this example, we choose $|\psi_1\rangle = |\psi_f\rangle$, $|\psi_2\rangle = e_1$. Then the Schmidt orthogonalization is performed. Suppose the orthogonalized vectors are $|s_1\rangle$ and $|s_2\rangle$, where $|s_1\rangle=|\psi_f\rangle$. According to (12), $P=p_1|s_1\rangle\langle s_1|+p_2|s_2\rangle\langle s_2|$ holds. The state except target state in $\mathcal{R}_2$ is $\rho_s=|s_2\rangle\langle s_2|$. To ensure the convergence, (9) is satisfied: $p_1<p_1 tr(\rho_f\rho_0)+p_2 tr(\rho_s\rho_0)<p_2$, then it gets $p_2>\max\{p_1*(1-tr(\rho_f\rho_0))/tr(\rho_2\rho_0),p_1\}$. According to Eq. (16), one gets $p_2=g_2*p_1$, where $g_2>\max\left\{\dfrac{1-tr(\rho_f\rho_0)}{tr(\rho_2\rho_0)},1\right\}$. If $p_1=0.2, g_2=10$ is selected, then $P=\begin{bmatrix}1.775 & -0.595\\-0.595 & 0.425\end{bmatrix}$. We denote $P$ with its Bloch vector as $\vec{P}=(x_1,x_2,x_3)=(-1.19,0,1.35)$. In fact, if $P$ is chosen as a real Hermitian matrix, one always gets $x_2=0$. According to Lemma 1, $\rho$ is expressed with Bloch vector as $\rho=\dfrac{1}{2}I+\sum_{i=1}^{3}\lambda_i\sigma_i$, then one gets $A(\vec{P})=\begin{pmatrix}0 & -x_3 & 0\\x_3 & 0 & -x_1\\0 & x_1 & 0\end{pmatrix}$ and $rank\tilde{A}(\vec{P})=2$, which satisfied the condition of lemma 1.

(2) Illustrations of system simulation experiments

The control gain in (6) is selected as $k=0.1$. The control law (6) with $P$ designed in Sec. 4 is applied to the system (2). Simulation experimental results are showed in Fig.1 in which Fig.1 (a) shows the state transferring process during the time $t\in[0,50]$. Fig.1 (b) is the control field. Red circle denotes the initial state of controlled system and blue circle is the target state; red line is the trajectory of the controlled state from initial point to its final point and the arrow indicates its transferring direction.

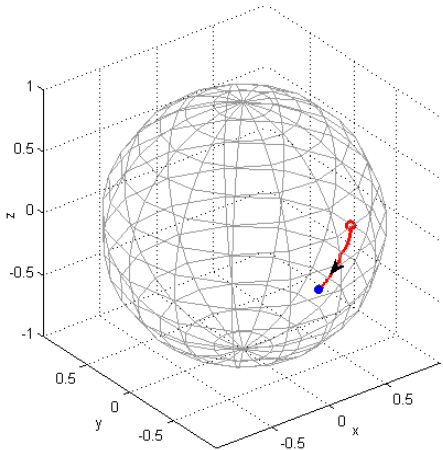
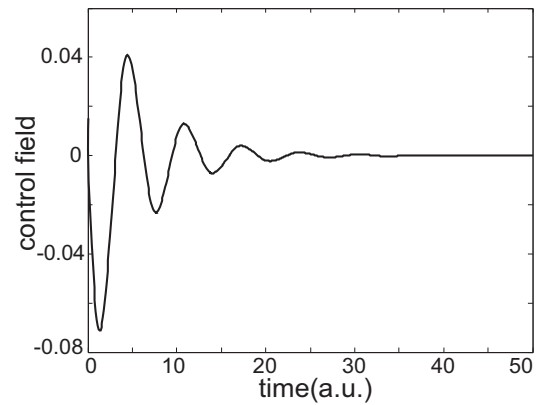

Fig. 1(a) Evolutionary trajectory of (2) during $t\in[0,50]$ a.u.        Fig. 1(b) Control field

To illustrate better the effectiveness of control strategy proposed in this paper, the control field Fig.1(b) is applied to the origianl system (1a). The tracking results are showed in Fig. 2, where red dashed line is the evolution curve of controlled state in (1a) and blue solid line is the one of target state in target system (1b); red circle and blue circle indicate the initial location at the current period of the controlled state and the target state, respectively; the arrow indicates the direction. The evolution trajectory during the time $t \in [0,8]$ a.u. is showed in Fig.2(a). What one can see the controlled system is asymptotically convergent to the target system on the Bloch sphere. Fig.2 (b) is the state trajectory during the time $t \in [8,30]$ a.u. and Fig.2(c) is the magnified bottom view of Fig.2 (b). We had specially labeled the different locations with black box. It can be seen from Fig.2(c) that red circle overlapped blue one at $t=30$ (the upper right box), so the tracking is achieved at this moment. Since then, the controlled system will follow the target state in target orbit. All the three figures demonstrate integrallty how the system (1a) tracks the (1b). If the performance index $v = \|\hat{\rho}(t) - \hat{\rho}_f(t)\|^2 = tr\left((\hat{\rho} - \hat{\rho}_f)^+ (\hat{\rho} - \hat{\rho}_f)\right)$ is used to measure the tracking accuracy, then $v = 9.41*10^{-5}$ is reached at $t = 50$ a.u..

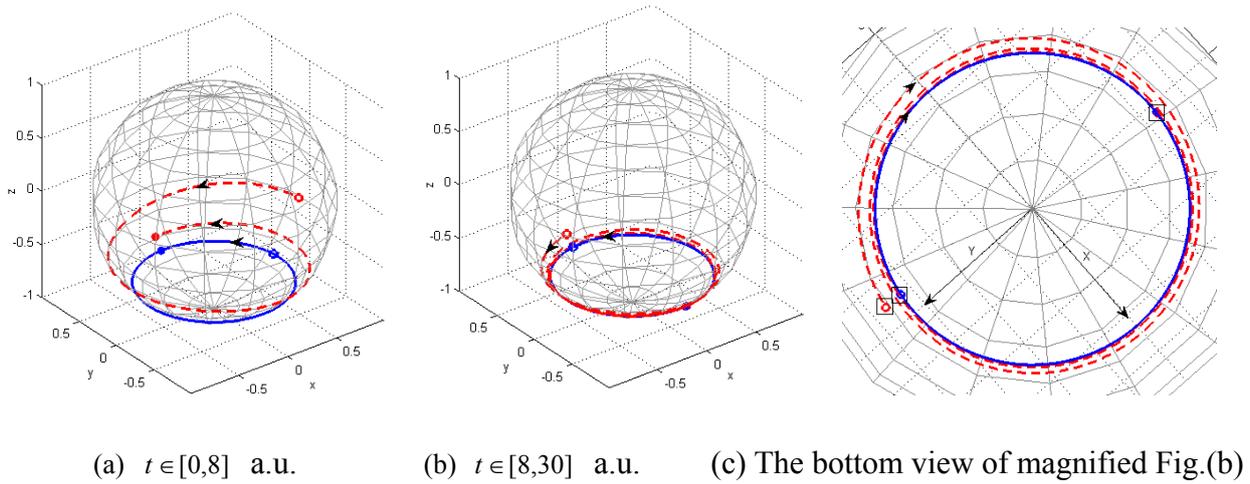

(a) $t \in [0,8]$ a.u.　　(b) $t \in [8,30]$ a.u.　　(c) The bottom view of magnified Fig.(b)

Fig.2 The trajectory tracking process of non-diagonal superposition target state

(3) Analysis of tracking properties

According to the weight coefficient $g_k$ defined in Eq.(16), the relationship between the two eigenvalues of $P$ is propotional. In this part, the experiment is used to demonstrate the effects of performance index produced by different cofficients $g_k$ in the case that other parameters remain unchanged. In the experiments, $g_2$ was selected as 3,6,12, respectively. The performance index is $v = \|\hat{\rho}(t) - \hat{\rho}_f(t)\|^2$. Fig.3 shows the influence of different $g_2$ on performance index, where the solid line is for $g_2=3$; the dot line and the dashed line are for $g_2=6$ and $g_2=12$, respectively. We also drawn the performance index in $100 < t < 150$ in Fig.3.

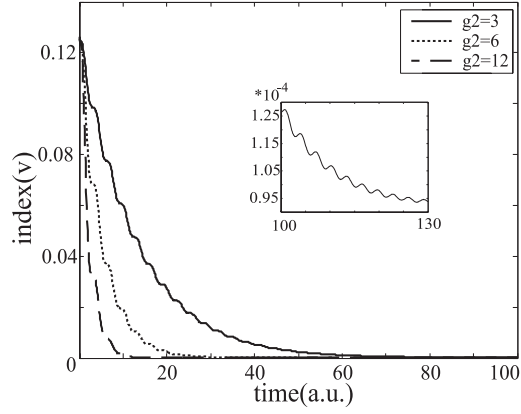

Fig. 3 Influence of different $g_2$ on performance index $v = \|\hat{\rho}(t) - \hat{\rho}_f(t)\|^2$

From Fig. 3 one can see that performance index $v = 9.26 \times 10^{-5}$ was achieved at $t$ = 15.1 a.u. for $g_2$ = 12; $v = 9.46 \times 10^{-5}$ was reached at $t$ = 39.6 a.u. for $g_2$ = 6; $v = 9.48 \times 10^{-5}$ was achieved at $t$ = 123.9 a.u. for $g_2$ = 3. It can be found that the coefficient $g_2$ determines the convergent rate. The larger $g_2$, the faster rate.

### 5.2 The case of non-stationary mixed initial target state

The initial states of (1a) and (1b) are $\hat{\rho}_0$=[0.45 0.274;0.274 0.55] and $\hat{\rho}_{f0}$=[0.762 -0.094;-0.094 0.238], respectively, which are both non-diagonal mixed-states. Because the purity of mixed-state is less than 1 and the initial state is unitary equivalently to the target state, the evolution of mixed-state is on a certain sphere inside the Bloch sphere. According to (2), a unitary transformation $U_T = U_f e^{iH_0 t}$ is firstly applied to (1a), where $U_f$=[0.985 -0.171;0.171 0.985]. And then Eq. (17) is obtained, where $\rho_f(0) = D_f = diag([0.778, 0.222])$ and $\rho(0)$=[0.361 0.241;0.241 0.639]. Then $P = diag([p_1, p_2])$ is constructed to meet (9) as: $p_1(D_f)_{11} + p_2(D_f)_{22} < p_1(\rho_0)_{11} + p_2(\rho_0)_{22} < p_1(D_f)_{22} + p_2(D_f)_{11}$, one gets $p_1 < p_2$. If we choose the $p_1$=0.1 and $g_2$=2, then $P = diag([0.1, 0.2])$ holds. Control gain in (6) is $k$=2. Fig. 4 is the state transferring process of non-diagonal mixed target state in which Fig.4 (a) is the evolutionary trajectory, red circle denotes the initial state of controlled system and blue circle is the target state; red line is the trajectory of the controlled state from initial point to its final point and the arrow indicates its direction. Fig.4 (b) is the control field.

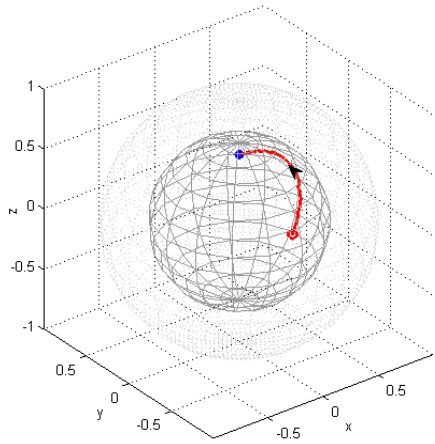 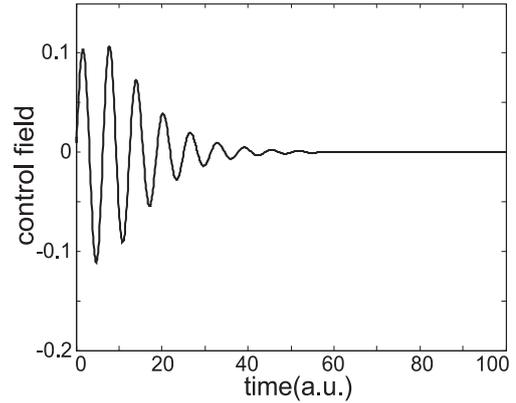

Fig. 4.1 state evolution　　　　　　　　　　　　　　　Fig. 4.2 control field

Fig. 4 The state transferring process of non-diagonal mixed target state

Similarly, the control field is applied to the original system (1), the trajectory tracking results are shown in Fig. 5 in which red line denotes the trajectory of the controlled state and blue denotes that of target state; Red and blue empty circles mean starting points of the controlled system and the target system, respectively and red and blue solid circles indicate final points; the black arrow indicates the tracking direction; the black boxes label the location of states. Fig.5 (a) shows the trajectory tracking process at $t \in [0, 14.2]$. Fig.5 (b) is the plane view of magnified Fig.5 (a). From Fig. 5 it can be seen that the controlled state is asymptotically convergent with respect to the target orbit from outside to inside. It entered the target orbit at $t=14.2$. At this time, however, owing to the target state (blue solid circle) is far from the controlled state (the red solid circle), the control law is non-zero. The system keeps the evolution. During the time $t \in [14.2, 60]$, the trajectory tracking is showed in Fig.5 (c) and Fig.5 (d) is the plan view of Fig.5 (c). In Fig. 5(d), the upper left box and the bottom right box label the locations of controlled state and target state at $t=14.2$, from this time, the controlled state firstly goes away from target orbit. Under control law, it gradually comes back to the target orbit from inside to outside and overlaps the target state at the upper right-hand location at $t=60$. And then both the controlled state and target state stay in target orbit and keep the same evolution. The performance $v=6.52*10^{-5}$ was arrived at $t=100$.

In a summary, for the initial target state, included diagonal ones and non-diagonal ones, the controlled system will converges to its target system under the control law (6) with $P$ designed as proposed in section 4.

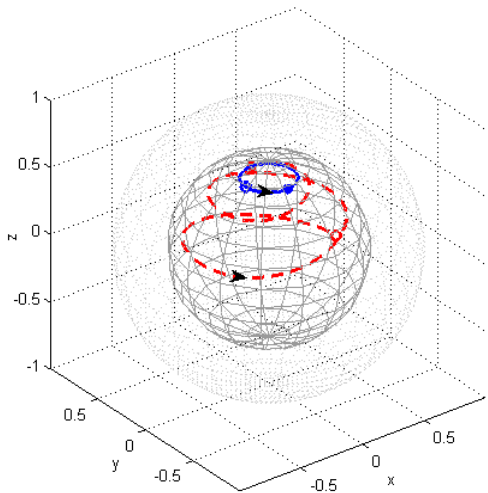 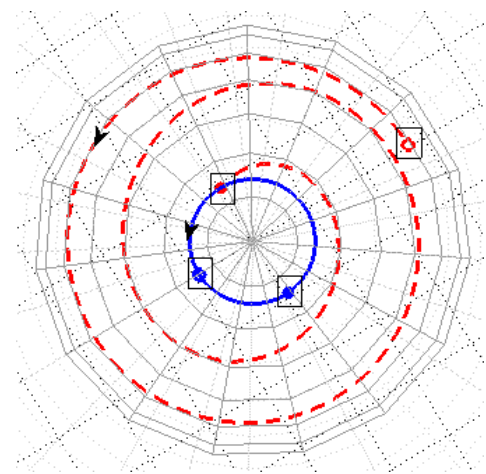

(a) The trajectory tracking of system (1) at $t \in [0, 14.2]$      (b) The plan view of magnified Fig.5 (a)

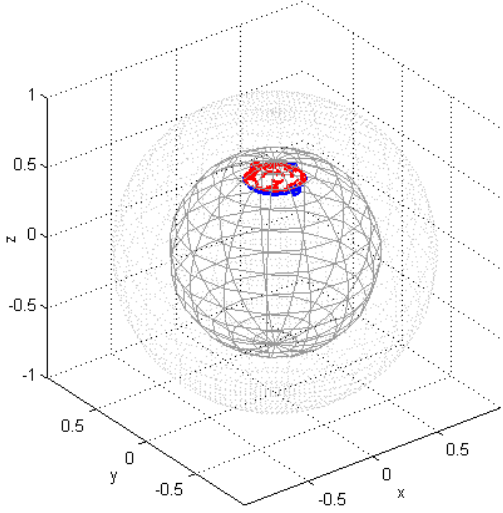 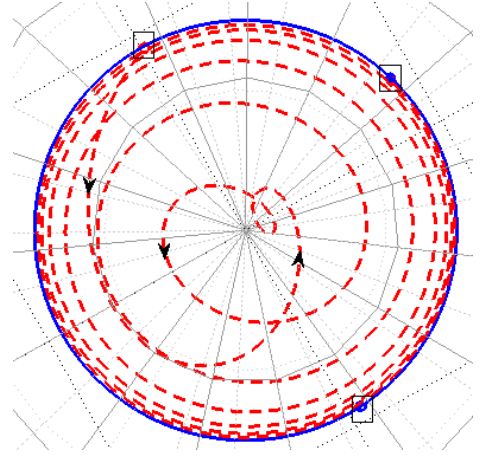

(c) The trajectory tracking of system (1) at $t \in [14.2, 60]$      (d) The plan view of magnified Fig.5 (c)

Fig. 5 The trajectory tracking process of non-diagonal mixed-state

## 6  Conclusion

In conclusion, we have proposed a control method for orbit tracking of free-evolutionary target quantum systems. The unitary transformation was used to change the tracking problem into the regulation one. For the convergence analysis in the process of state transferring, the target states are divided into diagonal state and non-diagonal state. For the former, based on the full study of eigen- target state, we studied the convergence of diagonal mixed-states in depth. The explicit convergence conditions of virtual mechanical quantity $P$ were obtained. For the non-diagonal target state, if the target state was a superposition state, a specific non-diagonal $P$ was designed to ensure the convergence. If the target state was a non-diagonal mixed-state, there must be a unitary transformation to change the Hermitian non-diagonal matrix into the diagonal matrix, and in such a case the convergence conditions could be obtained just as in the case of diagonal mixed-state. Based on the proposed control strategy, the orbit tracking is realized under the control of Lyapunov method. The control strategy proposed in this paper will be functioned under two assumptions mentioned, these assumptions are rather harsh. In practice, they may be non-existent. An implicit Lyapunov method has been proposed [18, 26], which can be used to design the suitable control laws for those systems which do not satisfy the assumptions.

**Appendix 1: The proof of Proposition 1.**

Proof:

The control Hamiltonian $H_m$ can be written as $H_m = H_{kl} = |k\rangle\langle l| + |l\rangle\langle k|$, then

$$H_{mt} = e^{iH_0 t}(|k\rangle\langle l| + |l\rangle\langle k|)e^{-iH_0 t} = e^{i\omega_{kl} t}|k\rangle\langle l| + e^{-i\omega_{kl} t}|l\rangle\langle k|, \text{ where } \omega_{kl} = \lambda_k - \lambda_l, \lambda_i (i=0,\cdots n) \text{ is the}$$

eigenvalue of $H_0$. Because $H_0$ is non-degenerate, $\omega_{kl} \neq 0$ holds.

The Eq. (8) becomes:

$$tr(H_m(t)[\rho_s,P]) = e^{iw_{kl}t}\langle l|[\rho_s,P]|k\rangle + e^{-iw_{kl}t}\langle k|[\rho_s,P]|l\rangle = 0$$

Then $\langle l|[\rho_s,P]|k\rangle=0$, viz. $([\rho_s,P])_{kl}=0$ holds. Let $A=[\rho_s,P]$, because of the hermiticity and positivity of $P$, $A$ is a skew Hermit matrix. Control Hamiltonian $H_m$ is full-connected, so $([\rho_s,P])_{kl}=0$ holds for all $k\neq l$, then $[\rho_s,P]=D$, where $D$ is a diagonal matrix.

Proposition 1 is proved.

**Appendix 2: The proof of Lemma 2.**

Proof:

$P$ is a diagonal matrix, one can get $\dot{v}(\rho_f)=0$ from Eq. (5).

$$\ddot{v}(\rho) = -i*\sum_m f_m\{tr(\dot{H}_{mt}[\rho,P]) + tr(H_{mt}[\dot{\rho},P])\}$$

$$\ddot{v}(\rho_f) = -\sum_m f_m^2 tr([H_{mt},\rho_f]*[P,H_{mt}])$$
$$= \sum_m f_m^2 tr([H_{mt},\rho_f]*[H_{mt},P])$$

Let $A=[H_{mt},\rho_f], B=[H_{mt},P]$, then $(A)_{ij}=(\lambda_j-\lambda_i)(H_{mt})_{ij}, (B)_{ij}=(p_j-p_i)(H_{mt})_{ij}$, so

$$tr(AB) = \sum_{i=1}^n\sum_{j=1}^n A_{ij}B_{ji} = \sum_{i=1}^n\sum_{j=1}^n (\lambda_j-\lambda_i)(p_i-p_j)(H_{mt})_{ij}^2 = -\sum_{i=1}^n\sum_{j=1}^n (\lambda_j-\lambda_i)(p_j-p_i)(H_{mt})_{ij}^2.$$

If $\rho_f$ is a stable state, then $\ddot{v}(\rho_f)>0$, one gets: $(\lambda_i-\lambda_j)(p_i-p_j)<0, \forall i\neq j$.

Let $\{\mu_1,\mu_2\cdots\mu_n\}$ be the spectrum of $\rho_f$ with $\mu_i$ arranged in a non-increasing order, viz $\mu_1<\mu_2<\cdots<\mu_n$. Then the corresponding $P$ is $P=diag(p_1,p_2,\cdots p_n)$ and $p_1>p_2>\cdots>p_n$ is got by the above description. One of the $P$ is $P=-\rho_f$, which is common in [14]. For any other states $\rho_s$ in limit set $\mathcal{R}$ can be obtained by $m$ times swapping arbitrary two elements of $\{\mu_1,\mu_2\cdots\mu_n\}$, let $bool = tr(P\rho_f)-tr(P\rho_s)$.

Suppose the spectrum from smallest to largest of target state is

$\{\mu_1, \mu_2, \cdots, \mu_i, \cdots, \mu_j, \cdots, \mu_k, \cdots \mu_n\}$, then:

$$i \leftrightarrow j: \quad bool = p_i(\mu_i - \mu_j) + p_j(\mu_j - \mu_i) = (p_i - p_j)(\mu_i - \mu_j) < 0$$

$i \leftrightarrow j, j \leftrightarrow k:$

$$\begin{aligned} bool &= p_i(\mu_i - \mu_j) + p_j(\mu_j - \mu_k) + p_k(\mu_k - \mu_i) \\ &= p_i(\mu_i - \mu_j) + p_j(\mu_j - \mu_i + \mu_i - \mu_k) + p_k(\mu_k - \mu_i) \\ &= (p_i - p_j)(\mu_i - \mu_j) + (p_j - p_k)(\mu_i - \mu_k) < 0 \end{aligned}$$

$i \leftrightarrow j, j \leftrightarrow k, k \leftrightarrow l:$

$$\begin{aligned} bool &= p_i(\mu_i - \mu_j) + p_j(\mu_j - \mu_k) + p_k(\mu_k - \mu_l) + p_l(\mu_l - \mu_i) \\ &= p_i(\mu_i - \mu_j) + p_j(\mu_j - \mu_i + \mu_i - \mu_k) + p_k(\mu_k - \mu_i + \mu_i - \mu_l) + p_l(\mu_l - \mu_i) \\ &= (p_i - p_j)(\mu_i - \mu_j) + (p_j - p_k)(\mu_i - \mu_k) + (p_k - p_l)(\mu_i - \mu_l) < 0 \end{aligned}$$

and so on. Finally, we get $v(\rho_f) < v(\rho_s)$. Lemma 2 is proved.

**ACKNOWLEDGEMENTS**

This work was supported in part by the National Key Basic Research Program under Grant No. 2011CBA00200, the Doctoral Fund of Ministry of Education of China under Grant No. 20103402110044, the National Science Foundation of China under Grant No. 61074050.